\documentclass[12pt]{article}
\input epsf.tex 
\def\beq{\begin{equation}} 
\def\eeq{\end{equation}} 
\def\lessim{\mathrel{\mathpalette\vereq<}} 
\def\tbar{\overline \theta} 
\def\Imtr#1{{\rm Im\: tr}\left[#1\right]}
\makeatletter 
\def\vereq#1#2{\lower3pt\vbox{\baselineskip1.5pt \lineskip1.5pt 
\ialign{$\m@th#1\hfill##\hfil$\crcr#2\crcr\sim\crcr}}} 
\makeatother 

\title{Solving the strong CP problem with supersymmetry 
 \footnote{Work supported by the Department of Energy,  
Contracts DE-AC03-76SF00515 (GH) and DE-AC02-76CH03000 (MS)}} 
\author{ 
        Gudrun Hiller\ $^{a}$\thanks{\tt ghiller@slac.stanford.edu}\ \  
        and Martin Schmaltz\ $^b$\thanks{\tt schmaltz@fnal.gov} \\ \\ \\ 
        \small \sl $^a$\ SLAC, Stanford University, Stanford, CA 94309\\ 
        \small \sl $^b$\ Theory Department, Fermilab, Batavia, IL 60510\\ \\ 
       } 
 
\begin{document} 
\baselineskip=17pt 
\pagestyle{plain} 
 
\begin{titlepage} 
\vskip-.4in 
\maketitle 
\begin{picture}(0,0)(0,0) 
\put(180,350){FERMILAB-Pub-01/037-T} 
\put(230,330){SLAC-PUB-8801} 
\end{picture} 
 
\begin{abstract} 
\leftskip-.6in 
\rightskip-.6in 
\vskip.4in 
We propose a new solution to the strong CP problem based on supersymmetric
non-renormalization theorems. CP is broken spontaneously and it's 
breaking is communicated to the MSSM by radiative corrections.
The strong CP phase is protected by a susy non-renormalization theorem
and remains exactly zero while loops can generate a large 
CKM phase from wave
function renormalization. We present a concrete model as an example but
stress that our framework is general. We also discuss constraints on susy
breaking and point out experimental signatures.

\end{abstract} 
\thispagestyle{empty} 
\setcounter{page}{0} 
\end{titlepage}

\section{Introduction} 
 
The strong CP problem \cite{reviews} has been a puzzle since the 70's 
when it was understood that the
$\theta$ parameter of QCD is physical. More recently, the strong CP problem has
sharpened because fits to $K$ and $B$ physics data now show 
that the unitarity triangle has three large angles \cite{pdg,recentckmfit}.
Thus superweak models are ruled out, and  CP violation in the CKM
matrix is large, the phase is order one.
The only other CP
violating parameter in the Standard Model, the strong CP 
phase, is 
experimentally bound
to be tiny \cite{strongnEDM}, most recent measurements of the electric 
dipole moment
of the neutron and $^{199}\!\!$~Hg imply $\tbar \lessim 10^{-10}$ 
\cite{tbarbounds}.
This asymmetry is especially puzzling since in the Standard Model
the CKM phase $\phi_{CKM}$ and the strong CP phase $\tbar$ have a
common 
origin.
Both of them come from the Yukawa couplings. The CKM matrix is the mismatch
between the basis in which the up and down quark Yukawa matrices 
are diagonal. In
the absence of fine-tuning, a large CKM phase implies large phases in
the Yukawa matrices. But if phases in the Yukawa matrices are large then the
bound on the strong CP phase
\beq
\tbar=\theta-\arg\ \det\ Y_u Y_d
\label{eq:tbar}
\eeq
implies fine-tuning of one part in $10^{10}$! Here and in the following we
assume real Higgs vacuum expectation values so that phases in quark masses
arise only from phases in Yukawa couplings.

The most popular proposed solutions to this problem are the axion \cite{axion},
a vanishing up-quark mass \cite{mu=0} and the Nelson-Barr mechanism 
\cite{nelsonbarr}.
For others see e.g.~\cite{others}.
For the axion solution $\tbar$ is promoted to be a field, the axion.
QCD dynamics generates a potential for the axion with a minimum at zero
as desired. The trouble with this solution is that experimental
searches 
for the axion have
found nothing and together with cosmological constraints have reduced
the 
allowed
parameter space to a small window \cite{pdg}.
A vanishing up-quark mass would nullify the strong CP
problem because it would render $\tbar$ unphysical. When $m_u=0$ then
the 
strong CP phase can
be removed from the Lagrangian by redefining the phase of the up quark field.
However, chiral perturbation theory disfavors 
$m_u=0$ \cite{leutwyler}.
While this possibility is still under debate,
the question will eventually be settled by lattice computations
\cite{CKN}. 
Finally, the
Nelson-Barr mechanism stipulates that CP is a good symmetry at high scales.
CP is broken spontaneously by a complex vev which is coupled to
the quarks in such a way that it induces complex mixing with heavy 
vector-like fermions.
By a clever choice of quark masses and Yukawa couplings, one can
arrange 
for a large CKM phase
and $\tbar = 0$. Loop corrections to $\tbar$ are dangerous in the 
Nelson-Barr scheme,
however they can be made sufficiently small by taking the coupling to the CP
violating vev small.

In this Letter we propose a new solution to the strong CP problem
which relies
on spontaneous CP violation and uses the non-renormalization theorems of
supersymmetry to ensure that $\tbar$ remains zero. 
Our basic framework assumes unbroken CP and susy at a high scale,
e.g.~the Planck scale. Therefore we can choose a basis in
which all coupling constants are real and $\theta=0$.
CP breaks spontaneously at the scale $M_{CP}$, and we assume that the MSSM
fields do not couple to complex vevs at tree level. Thus 
$\phi_{CKM}=\tbar=0$ at tree level.
However, a sufficiently large CKM phase is generated by loops if the 
CP violating sector 
(CPX) is strongly coupled to quarks. Naively one would expect that a
strong CP phase of order one is also generated. Happily, $\tbar$ is 
protected by a
non-renormalization theorem \cite{EFN,graesser}. Thus in susy
quantum loops can generate a large CKM phase while $\tbar$ remains 
exactly zero.
After susy breaking the non-renormalization
theorem no longer holds, and a small $\tbar$ is generated. We 
show that $\tbar$ remains sufficiently small if susy breaking occurs at 
low energies and is CP and flavor preserving.
Measurements of the superpartner spectrum, electric dipole
moments and CP and flavor physics in the $B$-system will test 
the predictions of our framework.

In the following section we discuss the
general framework in more detail. We give a specific model of CP violation
which exemplifies our mechanism in section 3. 
In section 4 we discuss susy breaking and 
in the fifth section we list model independent predictions. We 
conclude in section 6.

\section{CP phases from wave functions}

We now discuss our general scenario in more detail. 
Below the cut-off scale (for
example the Planck scale or the string scale) we assume that susy and
CP are good symmetries so that we can describe physics at this scale
by a local supersymmetric Lagrangian with real couplings and vanishing 
$\theta$-parameters. CP must be broken spontaneously at a lower scale $M_{CP}$ 
by the complex vev of one or several scalar fields $\Sigma$. 
In order to prevent a direct large contribution to $\tbar$ at this scale we
assume that there are no tree level superpotential couplings of
$\Sigma$ 
to the MSSM
or other colored fields. This could be enforced by a symmetry, or such 
couplings may be
suppressed for geometric reasons in extra dimensions \cite{nimaschmaltz}.
In order to communicate CP violation to the MSSM we assume that the CPX
sector couples to the quarks through messenger fields which couple to $\Sigma$.
There may also be arbitrary Kaehler potential couplings of $\Sigma$ to 
the MSSM. Such couplings are harmless, because the Kaehler potential
is real and
cannot contribute to $\tbar$ \cite{hilsch}.

This Lagrangian is renormalized at the loop level. In the following, 
it is convenient
to use the ``holomorphic'' renormalization scheme in which non-renormalization
theorems are manifest. Then both the superpotential and $\tbar$ are
not 
renormalized.
However the Kaehler
potential is renormalized, and if the CP messenger sector and it's 
couplings violate
flavor non-canonical complex kinetic terms for the quarks are induced.
The most general CP violating kinetic terms are $3\times3$ hermitian 
matrices $Z$
for each set of fields with identical gauge quantum numbers.
Because the $Z$'s are hermitian and positive definite we can write $Z^{-1}=T^2$
with a hermitian $T$ and change to the canonical basis
\beq
L \sim \int d^4\theta\ Z^{ij} \hat Q^\dagger_i \hat Q_j
   = \int d^4\theta\ \delta^{ij} Q_i^\dagger Q_j\quad {\rm where}\quad  Q=T^{-1}\hat Q \ .
\label{eq:wavefunc}
\eeq
Note that wave-function renormalization by the hermitian matrix leaves 
$\theta$
invariant~\footnote{This was pointed out to us by 
Holdom but has been known by others as
well \cite{EG,DGH,holdom}}, because $\theta$ shifts proportional to 
$\arg  \det \, T = 0$. In the new basis the Yukawa terms
are $Q^T Y_u U H_u$,  $Q^T Y_d D H_d$, where 
\beq
Y_d = T_q^T \hat Y_d T_d\quad {\rm and}\quad Y_u = T_q^T \hat Y_u T_u\ ,
\label{eq:yuks}
\eeq
and also
\beq 
\tbar \ = \ \theta-\arg\, \det\, Y_d Y_u \ 
= \ 0 - \arg\, \det\, \hat Y_d \hat Y_u \ = \ 0 \ , 
\eeq
reflecting the non-renormalization of $\tbar$.

However, even though $\tbar$ remains zero, the new Yukawa matrices are clearly
complex and a complex CKM matrix is generated. It is intuitive
(but somewhat tedious to show \cite{hilsch}) that a large CKM phase is
generated if the wave-function renormalization factors $T$ are not close
to the unit matrix. Thus if the CP violating dynamics breaks CP by order one
and is strongly coupled to the quarks then we obtain
\beq
\phi_{CKM} \sim O(1) \ ,\quad \tbar = 0\
\label{eq:tbar=0}
\eeq
as desired. It is also easy to show that all values for quark masses, 
mixing angles
and CKM phase can be generated in this way. To see this, pick as an example
$\hat Y_d=\hat Y_u=1=T_q\ ,\ T_u \propto {\rm diag}(m_u,m_c,m_t)\ ,\ 
T_d\propto V_{CKM} {\rm diag}(m_d,m_s,m_b) V_{CKM}^\dagger$.
Before discussing susy breaking and
corrections to eq.~(\ref{eq:tbar=0}) in section \ref{sec:susyx}
we give an explicit example for the CP violating sector. 

\section{An explicit model}
\label{sec:model}

In this section, we give one of many possible models as an example. This
model generates large wave function renormalizations only for the down quark
singlet which is sufficient to obtain a large CKM phase. Other models with
$SU(5)$ or $SO(10)$ unification and messenger fields in full representations
of the GUT group can also be built and may be more attractive/predictive.
In addition to the usual
MSSM superpotential and canonical kinetic terms we assume the following
superpotential at the high scale, $M_{Pl}$
\beq
W=r_{ij} D_i  F_j T + M_{CP} T \overline T +
\Sigma _{ij} F_i \overline F_j \ .
\eeq
Here $T$ and $\overline T$ are a vector-like, fourth-generation down quark
singlet with real mass $M_{CP}$, $F$ and $\overline F$ are three vector-like
SM singlets which obtain complex masses from their coupling to the complex vev
$\| \Sigma \| \, \sim \, M_{CP}$
and we have absorbed a coupling constant into the definition of $\Sigma$.
\begin{figure}[htb]
\vskip 0.0truein
\centerline{\epsfysize=1.8in
{\epsffile{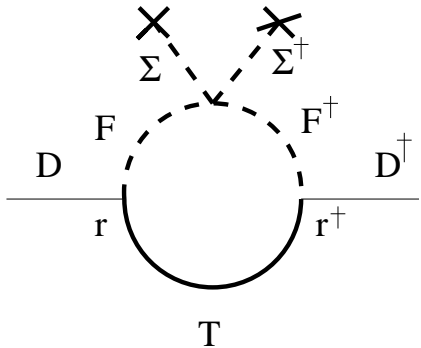}}}
\vskip 0.0truein
\caption[]{\it A diagram contributing to the down quark kinetic term $Z_d$.}
\label{fig:legs}
\end{figure}
%
%
%

Integrating out the massive fields at the scale $M_{CP}$ sets the
above superpotential to zero, the MSSM superpotential remains unchanged
in the ``holomorphic'' renormalization scheme, and the one-loop
diagram of Figure 1 renormalizes the wave function of 
the right handed down quarks
\beq
\label{eq:downrenorm}
\delta Z_d \ \sim\ {r^\dagger \Sigma^\dagger \Sigma r \over 16 \pi^2 M_{CP}^2}\ \times\ {\rm Log}\ .
\eeq
Note that the coupling constant $r$ needs to be strong ($\sim 4\pi$) in
order for $Z_d$ and $T_d$ not to be close to the unit matrix. The strong
coupling renders this one-loop calculation of $Z$ unreliable,
and we simply parameterize the full result by $T_d$.%
\footnote{Maintaining such a large Yukawa coupling at the scale $M_{CP}<M_{Pl}$
requires new strong gauge interactions for the $T$'s and $F$'s. We have checked
that quantum corrections involving these new gauge interactions do not 
spoil our
mechanism, neither perturbatively nor non-perturbatively \cite{hilsch}.}
It is important that
there are no true vertex renormalizations of $Y_d$ due to the
non-renormalization theorem. This should highlight why susy is
crucial for our approach.
In a non-supersymmetric theory one-particle-irreducible
vertex corrections do arise at some order, and
because of the strong coupling
diagrams with arbitrarily many loops are dangerous.

At the renormalizable level, the wave-function renormalization
factor $T_d$ is the only parameter which remains from the CP violating 
dynamics at scales below $M_{CP}$. Higher dimensional operators suppressed
by $M_{CP}$ are also generated. In the presence of susy breaking they lead
to important corrections to $\tbar$ as we discuss in Section \ref{sec:susyx}.

Finally, we note that we can relax the condition that
$\Sigma$ not couple at all to MSSM fields in the superpotential to a less
stringent constraint by allowing couplings at the non-renormalizable level.
Operators such as 
\beq
{\rm tr}\ \left({\Sigma \over M_{Pl}}\right)^n\ W_\alpha W^\alpha
\eeq
result in contributions to $\tbar \sim \left(M_{CP}/ M_{Pl}\right)^n$
which implies $M_{CP}/M_{Pl} \lessim 10^{-10/n}$. The existence of
such operators
is model dependent, and the upper bound on $M_{CP}$ is
therefore not mandatory.

\section{Supersymmetry breaking}
\label{sec:susyx}

Since susy is broken in nature we need to check that our
mechanism for protecting $\tbar$ from radiative corrections remains stable
after susy breaking. In this Letter we only discuss this
topic very briefly, a more detailed discussion is in our longer paper 
\cite{hilsch},
and many of the results of this section can also be found in
\cite{DGH,gmsbreview,barr,DKL}.

The non-renormalization theorems
are violated in the presence of susy breaking, and $\tbar$ is renormalized.
One contribution to $\tbar$ that is always there is the well-known
heavily GIM suppressed and therefore finite SM contribution which arises at
four-loops \cite{EG} from the ``cheburashka'' diagram \cite{khriplovich} and
gives $\delta \tbar \simeq 10^{-19}$. The diagram is dominated by loop momenta
near the QCD scale and is therefore independent of the mechanism of
susy breaking.

\begin{figure}[htb]
\vskip 0.0truein
\centerline{\epsfysize=1.5in
{\epsffile{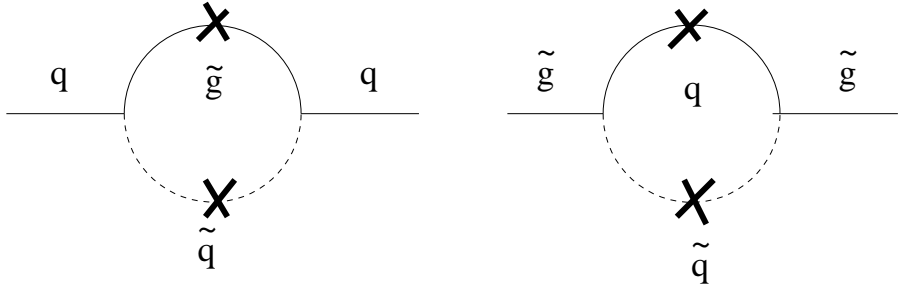}}}
\vskip 0.0truein
\caption[]{\it Lowest order susy diagrams contributing to $\tbar$. 
A cross denotes a LR mass insertion. }
\label{fig:loop}
\end{figure}

However, there are also new contributions which are specific to softly broken
susy theories. In the MSSM the expression for $\tbar$ must be generalized
to include the phase of the gluino mass $m_{\tilde g}$
\beq
\tbar=\theta-\arg\, \det\, Y_u Y_d - 3\, \arg\, m_{\tilde g} \ .
\label{eq:susytbar}
\eeq
Thus the gluino mass must be real to one part in $10^{10}$.
The same bound applies to the b-term, because a complex b leads to
complex Higgs vevs.
The phases of the remaining flavor-universal MSSM parameters
are also tightly constrained $\sim 10^{-8}$,
because they induce contributions to  $\tbar$
from the one-loop  diagrams in Figure 2.

But even if all phases in soft terms vanish the diagrams of Figure 2 can
still generate large contributions to $\tbar$ because they involve
the complex MSSM Yukawa couplings. The diagrams give expressions
such as $\Imtr{Y^\dagger A}$ and renormalize $\tbar$ unless $A$ is
either zero or else proportional to $Y$. The most natural way to suppress
these contributions is to assume that susy breaking is universal and 
proportional
\beq
\label{eq:universality}
m_{\tilde{u}}^2 \sim m_{\tilde{d}}^2 \sim m_{\tilde{q}}^2 
\propto 1\ , \quad A_{u/d} \propto Y_{u/d} \ .
\eeq
Then $\Imtr{Y^\dagger A}=0$, and all other similar traces vanish to 12th
order in an expansion of Yukawa couplings \cite{hilsch,DGH}, and therefore
contributions from susy breaking to $\tbar$ are negligibly small.

Very near universality and proportionality as 
in eq.~(\ref{eq:universality}) are 
required at
the weak scale. A susy breaking and communication mechanism which
accomplishes this is gauge mediation. This is discussed in more
detail in \cite{hilsch}, where we find that our mechanism works very
naturally with gauge mediation \cite{GMSB}.
Other susy breaking scenarios which are compatible
are anomaly mediation \cite{AMSB} and gaugino mediation \cite{GinoMSB}.
The scenario cannot, however, be combined with minimal supergravity.
The reason being that even if in mSUGRA soft masses are assumed
to be universal at the Planck scale they are
strongly renormalized by the strong CP and flavor violating dynamics
at 
$M_{CP}$ and
become completely non-degenerate. Inserting these non-degenerate soft 
masses into
the diagrams of Figure 2 gives disastrously large contributions to $\tbar$.
This argument implies quite generally (anomaly mediation is an
important 
exception)
that susy breaking needs to be communicated to the MSSM at a scale well
below $M_{CP}$. 

As the messenger scales of gauge mediation and CP violation get near each
other corrections to $\tbar$ proportional to $\left(M_{susy}/M_{CP}\right)^2$
arise and give the bound $M_{susy}/M_{CP}\lessim 10^{-3}$ \cite{hilsch} from
requiring $\tbar \lessim 10^{-10}$.
Note that if this bound is saturated then neutron dipole moment measurements
should find non-vanishing results soon.
Using the lowest possible messenger scale in gauge mediation 
$M_{susy} \sim 10^4$ GeV we find $M_{CP}>10^7$ GeV.
Therefore the particles of the CP violating sector cannot be produced
at existing or planned accelerators. However there are 
several
indirect predictions from our scenario which allow it to be tested. 
We discuss them in the next section.

\section{Predictions}
\label{sec:predictions}

Even though the CPX dynamics of our solution to the strong CP 
problem necessarily hides at short distances there are several testable
consequences. 
We predict \cite{hilsch}:

\begin{enumerate}
\item Supersymmetry
\item Minimal flavor violation, i.e., no new flavor
violation beyond the Yukawa couplings with the well-known implications for
$B$-physics \cite{GGG,buras2}.
\item No new CP violation beyond the SM, in particular
no new CP violation in the $B$-system. For example 
$\sin 2 \beta$ is large as in the SM \cite{buras2}.
\item Almost degenerate first and second generation scalars of each 
gauge quantum number. The splittings are proportional to the square of the
corresponding Yukawa couplings and give 
$\triangle m < 1$ GeV which should be measurable at a
linear collider. 
We stress that this degeneracy has to hold independent 
of the susy breaking mechanism.

\item $\tbar$ is predicted to lie between the current experimental bound of
$10^{-10}$ and $10^{-19}$ depending on the ratios of scales 
$M_{susy}/M_{CP}$ and
$M_{CP}/M_{Pl}$. If we are lucky the corresponding hadron electric 
dipole moments will
be measured soon \cite{golublamoreaux}. 
Lepton dipole moments are expected to be much smaller.

\end{enumerate}

\section{Conclusions}

The strong CP problem has recently become more urgent because experimental data
strongly favor a CKM phase of order one, 10 orders of magnitude larger than the
upper bound on the strong CP phase. This represents a puzzle because
both 
appear
to arise from Yukawa couplings in the SM. In this Letter we propose a new
solution where CP is broken spontaneously and mediated to the
SM by radiative corrections. Obtaining a large CKM phase requires the radiative
corrections to be large, forcing us to consider strongly coupled
models. 
Whereas such models are very difficult if not impossible to build without susy
we have argued
that the non-renormalization theorems of susy make such a solution to
the strong
CP problem very natural. The CKM phase gets $O(1)$ contributions from 
renormalization
whereas the strong CP phase remains exactly zero in the supersymmetric
limit.
Our picture requires flavor-universal susy breaking and mediation and
is compatible with
gauge-, anomaly-, and gaugino-mediation but not compatible with
minimal supergravity.
We presented an explicit model for the CP messenger sector, but we
stress that the framework is much more general because it is based on 
model-independent
non-renormalization theorems. It would be interesting to build
complete GUT models
based on our framework, possibly with flavor and CP violation
originating from the
same strongly coupled dynamics. A promising avenue to pursue is to combine
our framework with the models of Nelson and Strassler \cite{NS}.

\section{Acknowledgements}
 
We thank Darwin Chang, Sekhar Chivukula, Andy Cohen,
Lance Dixon, Gia Dvali, Howard Georgi, David E. Kaplan, Ann Nelson,
Michael Peskin, Joao Silva, Witek Skiba, Matt Strassler, 
Raman Sundrum, Neal Weiner and
especially Bob Holdom for useful discussions. We also thank Mary K. Gaillard,
Lawrence Hall and Iosif Khriplovich for communication on the
renormalization of $\theta$ in the SM.
GH thanks the theory department at Fermilab for
hospitality during her
stay, where part of this work has been done.


\begin{thebibliography}{99} 
 
\bibitem{reviews}
M.~Dine,
hep-ph/0011376; 
R.~D.~Peccei,
DESY-88-109
{\it  in Jarlskog, C. (Ed.): CP Violation 503-551}.

\bibitem{pdg}
D.~E.~Groom {\it et al.}  [Particle Data Group Collaboration],
Eur.\ Phys.\ J.\ C {\bf 15}, 1 (2000).

\bibitem{recentckmfit}
A.~Ali and D.~London,
Eur.\ Phys.\ J.\ C {\bf 18}, 665 (2001)
[hep-ph/0012155];
M.~Ciuchini {\it et al.},
hep-ph/0012308;
S.~Mele,
hep-ph/0103040.

\bibitem{strongnEDM}
R.~J.~Crewther, P.~Di Vecchia, G.~Veneziano and E.
Phys.\ Lett.\ B {\bf 88}, 123 (1979), Erratum-{\it ibid.}\ B {\bf 91}, 487 (1980); 
V.~Baluni,
Phys.\ Rev.\ D {\bf 19}, 2227 (1979);
M.~Pospelov and A.~Ritz,
Nucl.\ Phys.\ B {\bf 573}, 177 (2000)
[hep-ph/9908508].

\bibitem{tbarbounds}
P.~G.~Harris {\it et al.},
Phys.\ Rev.\ Lett.\ {\bf 82}, 904 (1999);
M.~V.~Romalis, W.~C.~Griffith and E.~N.~Fortson,
Phys.\ Rev.\ Lett.\ {\bf 86}, 2505 (2001)
[hep-ex/0012001].

\bibitem{axion}
R.~D.~Peccei and H.~R.~Quinn,
Phys.\ Rev.\ Lett.\ {\bf 38}, 1440 (1977);
Phys.\ Rev.\ D {\bf 16}, 1791 (1977).
S.~Weinberg,
Phys.\ Rev.\ Lett.\ {\bf 40}, 223 (1978);
F.~Wilczek,
Phys.\ Rev.\ Lett.\ {\bf 40}, 279 (1978).

\bibitem{mu=0}
G.~'t Hooft,
Phys.\ Rev.\ Lett.\ {\bf 37}, 8 (1976);

\bibitem{nelsonbarr}
A.~Nelson,
Phys.\ Lett.\ B {\bf 136}, 387 (1984);
S.~M.~Barr,
Phys.\ Rev.\ Lett.\ {\bf 53}, 329 (1984).

\bibitem{others}
M.~A.~Beg and H.-S.~Tsao,
Phys.\ Rev.\ Lett.\  {\bf 41}, 278 (1978);
R.~N.~Mohapatra and G.~Senjanovic,
Phys.\ Lett.\ B {\bf 79}, 283 (1978);
R.~N.~Mohapatra and A.~Rasin,
Phys.\ Rev.\ Lett.\  {\bf 76}, 3490 (1996)
[hep-ph/9511391];
H.-C.~Cheng and D.~E.~Kaplan,
hep-ph/0103346;
C.~Hamzaoui and M.~Pospelov,
hep-ph/0105270.

\bibitem{leutwyler}
H.~Leutwyler,
Nucl.\ Phys.\ B {\bf 337}, 108 (1990).

\bibitem{CKN}
A.~G.~Cohen, D.~B.~Kaplan and A.~E.~Nelson,
JHEP {\bf 9911}, 027 (1999)
[hep-lat/9909091].

\bibitem{EFN}
J.~Ellis, S.~Ferrara and D.~V.~Nanopoulos,
Phys.\ Lett.\ B {\bf 114}, 231 (1982).

\bibitem{graesser}
M.~Graesser and B.~Morariu,
Phys.\ Lett.\ B {\bf 429}, 313 (1998)
[hep-th/9711054].

\bibitem{nimaschmaltz}
N.~Arkani-Hamed and M.~Schmaltz,
Phys.\ Rev.\ D {\bf 61}, 033005 (2000)
[hep-ph/9903417].

\bibitem{hilsch}
G.~Hiller and M.~Schmaltz, {\it in preparation}.

\bibitem{EG}
J.~Ellis and M.~K.~Gaillard,
Nucl.\ Phys.\ B {\bf 150}, 141 (1979).
 

\bibitem{DGH}
M.~Dugan, B.~Grinstein and L.~Hall,
Nucl.\ Phys.\ B {\bf 255}, 413 (1985).

\bibitem{holdom}
B.~Holdom,
Phys.\ Rev.\ D {\bf 61}, 011702 (2000)
[hep-ph/9907361].


\bibitem{gmsbreview}
G.~F.~Giudice and R.~Rattazzi,
Phys.\ Rept.\  {\bf 322}, 419 (1999)
[hep-ph/9801271].

\bibitem{barr}
S.~M.~Barr and G.~Segre,
Phys.\ Rev.\ D {\bf 48}, 302 (1993);
%
S.~M.~Barr,
Phys.\ Rev.\ D {\bf 56}, 1475 (1997)
[hep-ph/9612396].



\bibitem{DKL}
M.~Dine, R.~G.~Leigh and A.~Kagan,
Phys.\ Rev.\ D {\bf 48}, 2214 (1993)
[hep-ph/9303296].


\bibitem{khriplovich}
I.~B.~Khriplovich,
Phys.\ Lett.\ B {\bf 173}, 193 (1986).

\bibitem{GMSB}
M.~Dine and A.~E.~Nelson,
Phys.\ Rev.\ D {\bf 48}, 1277 (1993)
[hep-ph/9303230];
M.~Dine, A.~E.~Nelson and Y.~Shirman,
Phys.\ Rev.\ D {\bf 51}, 1362 (1995)
[hep-ph/9408384];
M.~Dine, A.~E.~Nelson, Y.~Nir and Y.~Shirman,
Phys.\ Rev.\ D {\bf 53}, 2658 (1996)
[hep-ph/9507378].

\bibitem{AMSB}
L.~Randall and R.~Sundrum,
Nucl.\ Phys.\ B {\bf 557}, 79 (1999)
[hep-th/9810155];
G.~F.~Giudice, M.~A.~Luty, H.~Murayama and R.~Rattazzi,
JHEP {\bf 9812}, 027 (1998)
[hep-ph/9810442].

\bibitem{GinoMSB}
D.~E.~Kaplan, G.~D.~Kribs and M.~Schmaltz,
Phys.\ Rev.\ D {\bf 62}, 035010 (2000)
[hep-ph/9911293];
Z.~Chacko, M.~A.~Luty, A.~E.~Nelson and E.~Ponton,
JHEP {\bf 0001}, 003 (2000)
[hep-ph/9911323];
M.~Schmaltz and W.~Skiba,
Phys.\ Rev.\ D {\bf 62}, 095005 (2000)
[hep-ph/0001172].


\bibitem{GGG}
G.~Buchalla, G.~Hiller and G.~Isidori,
Phys.\ Rev.\ D {\bf 63}, 014015 (2001)
[hep-ph/0006136].

\bibitem{buras2}
A.~J.~Buras and R.~Buras,
Phys.\ Lett.\ B {\bf 501}, 223 (2001)
[hep-ph/0008273].


\bibitem{golublamoreaux}
R.~Golub and K.~Lamoreaux,
Phys.\ Rept.\ {\bf 237}, 1 (1994).



\bibitem{NS}
A.~E.~Nelson and M.~J.~Strassler,
JHEP{\bf 0009}, 030 (2000)
[hep-ph/0006251];
hep-ph/0104051.

\end{thebibliography}
\end{document}